# Quantitative Analysis and Efficiency Study of PSD Methods for a LaBr$_3$:Ce Detector

Ming Zeng [a,b], Jirong Cang [a,b], Zhi Zeng [a,b,*], Xiaoguang Yue[a,b], Jianping Cheng [a,b], Yinong Liu[a,b], Hao Ma[a,b] and Junli Li[a,b]

[a] *Key Laboratory of Particle & Radiation Imaging(Tsinghua University), Ministry of Education, China*

[b] *Department of Engineering Physics, Tsinghua University, Beijing 100084, China*

**Abstract**

The LaBr$_3$:Ce scintillator has been widely studied for nuclear spectroscopy because of its optimal energy resolution (<3% @ 662 keV) and time resolution (~300 ps). Despite these promising properties, the intrinsic radiation background of LaBr3:Ce is a critical issue, and pulse shape discrimination (PSD) has been shown to be an efficient potential method to suppress the alpha background from the $^{227}$Ac. In this paper, the charge comparison method (CCM) for alpha and gamma discrimination in LaBr3:Ce is quantitatively analysed and compared with two other typical PSD methods using digital pulse processing. The algorithm parameters and discrimination efficiency are calculated for each method. Moreover, for the CCM, the correlation between the CCM feature value distribution and the total charge (energy) is studied, and a fitting equation for the correlation is inferred and experimentally verified. Using the equations, an energy-dependent threshold can be chosen to optimize the discrimination efficiency. Additionally, the experimental results show a potential application in low-activity high-energy γ measurement by suppressing the alpha background.

## 1 Introduction

The LaBr$_3$:Ce scintillator has been widely studied for nuclear spectroscopy because of its optimal energy resolution, good efficiency, and excellent time resolution. However, $^{227}$Ac exists as a radioactive contamination because of the production technology, contributing to the intrinsic alpha background with energy above 1.6 MeV [1], which limits the application of LaBr$_3$:Ce to low-activity gamma measurement in this region. Thus, studying the discrimination of alpha and gamma events is of significance to the high-energy γ experiments and artificial radioactivity measurement.

In past years, pulse shape discrimination (PSD) methods for n-γ discrimination in organic scintillation detectors [2,3,4] and for α-γ-nuclear recoil discrimination in CsI(Tl) [5] have been well studied. Recently, a digital PSD method has been used with fast digitizers to study the PSD parameters in detail, as well as to identify the different components in the scintillation light and its underlying physics [3,4]. Concerning the LaBr$_3$:Ce crystal, the expected difference in pulse shape between alpha and gamma events is very small, as LaBr$_3$:Ce seems to lack different components within the scintillation light. Hoel first analysed the pulse shape difference in LaBr$_3$:Ce and concluded that PSD is not applicable, due to the small difference[6]. Later, however, Crespi achieved preliminary PSD results by using the Charge Comparison Method (CCM) with a much faster digitizer (2 Gsps) and showed the potential for suppressing the intrinsic alpha background[7]. Overall, the quantitative study of PSD for LaBr$_3$:Ce remains limited at present. With such research, it would be possible to optimize the discrimination efficiency and identify the reason for the pulse shape difference in LaBr$_3$:Ce or the existence of different components in the scintillation light.

* Corresponding author, *E-mail*: zengzhi@tsinghua.edu.cn

In this paper, digital CCM is first compared quantitatively with two other typical PSD methods, with parameter optimization for each method. The effectiveness of α-γ discrimination in the LaBr$_3$:Ce crystal is evaluated. In addition, it is shown that the discrimination efficiency varies with energy. Considering digital CCM, the correlation between the distribution of the CCM feature value and the total charge (energy) is also studied as well. A fitting equation for the correlation is inferred and verified using the experiment data.

## 2  Experiment

In this research, a 2x2 inch cylindrically shaped LaBr$_3$:Ce detector was used, which is commercially available from Saint-Gobain. The photomultiplier tube coupled to the crystal was a Hamamatsu R6233-100. For the full digitization of pulse shapes, a 2.5 Gsps 12-bit LeCroy Oscilloscope (HDO6104) was used to digitize the raw PMT output.

Very low-activity $^{137}$Cs and $^{22}$Na sources were used in this research (see Fig. 1) to generate sufficient γ events, while 5 cm lead shielding was used to reduce the influence of environmental radioactivity. Besides, the intrinsic radioactivity of LaBr3:Ce can be found in the Fig. 1, including the 789 keV gamma associated with the 256 keV beta, the 1436 keV gamma associated with the X-rays and the alpha between 1.8 - 2.5 MeV. For the comparison of PSD methods, the photo-peak of $^{137}$Cs at 662 keV was chosen as a typical γ event for the study, whilst the alpha particle events were from the intrinsic radioactivity within the energy ranges from 1.8 MeV to 2.5 MeV. Furthermore, the discrimination efficiency at different energies was studied using the full pulse shape data set (full energy range).

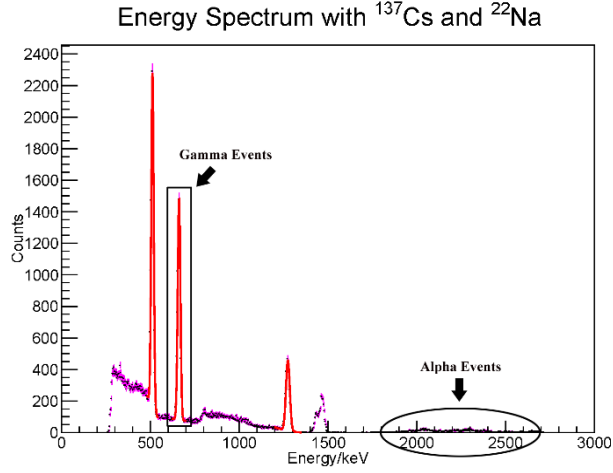

Fig. 1. Energy spectrum of nuclides used in this experiment and the particular region selected for typical alpha and gamma events

## 3  PSD method comparison and optimization of parameters

First, the CCM was compared quantitatively with 2 other typical PSD methods, the MTM and the GAMA methods, applied to α-γ discrimination for the same data set using digital pulse processing. The discrimination efficiency of each method was calculated.

### 3.1  Charge Comparison Method (CCM)

The Charge Comparison Method is a classical pulse shape discrimination method, based on a comparison of two different integrals of the current pulse signal. Normally, the long integral is the whole input of the current pulse. With digital CCM, the short integral can be chosen to correspond to the interval in which the difference between the α signals and the γ signals is most significant. As illustrated in Fig. 2, pulses were aligned according to their maximum at 40 ns. Based on comparing the area-normalized, averaged pulse shape of 2000 events of each type, the ratio between the difference of the two pulse shapes and the α one was calculated and illustrated in the inset of Fig.2. And the optimized short integral interval 25.2~68.0 ns was chosen using the following two rules:

1. The difference between alpha and gamma is maximized.
2. The amplitude value at the two boundaries is similar to minimize the uncertainty of CCM caused by the time jitter of the pulse alignment.

The CCM feature value is defined as the discrimination feature:

$$\text{CCM} = \frac{Q_p}{Q_t} = \frac{\int_{t_1}^{t_2} i(\text{t})dt}{\int_{total} i(\text{t})\,dt} \qquad (1)$$

where $t_1$ and $t_2$ are the lower and upper boundaries of the short interval. The discrimination efficiency is shown in Fig. 3.

### 3.2  Mean Time Method (MTM)

As discussed in [5], for lower-energy event discrimination with the CsI(Tl) scintillator detector,

the Mean Time Method is superior to conventional CCM.

The MTM feature value is defined as follows:

$$<t> = \frac{\sum_i (A_i t_i)}{\sum_i A_i} \qquad (2)$$

where $A_i$ is the Flash ADC (FADC) amplitude at time-bin $t_i$.

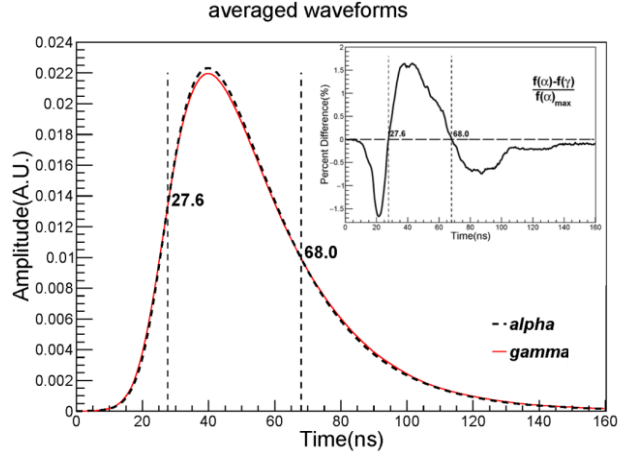

Fig. 2. The signal line-shapes measured for α-particles and γ-rays in LaBr3:Ce detector. Each pulse is the result of an average of 2000 pulses, normalized by area and aligned by the maximum (assume the time of the maximum as 40ns), with the energy region mentioned above. In the inset the ratio between the difference of the two pulse shapes and the α one is displayed.

For optimization of MTM, a good time alignment is essential to compare the pulse shapes. In this research, digital constant fraction discrimination (dCFD) is used to align these pulse shapes. Experimentally, the fraction was chosen as 20%, which gives good discrimination efficiency. The result is illustrated in Fig. 3.

## 3.3 **Gamma-Alpha Model Analysis (GAMA)**

With GAMA, the alpha and gamma pulse shapes were modelled prior to the analysis of the experimental data. The models were extracted from the average of a set of several thousand known alpha and gamma pulses (i.e., Fig. 1). Each unknown pulse was compared with the modelled ones to find the better more using Eq.3.

We define $p_u$, $m_\gamma$ and $m_\alpha$, normalized by total charge, as the unknown pulse, model gamma pulse and model alpha pulse, respectively:

$$\chi_\gamma^2 = \sum_{i=1}^{n} \frac{(p_u(i) - m_\gamma(i))^2}{m_\gamma(i)}$$

$$\chi_\alpha^2 = \sum_{i=1}^{n} \frac{(p_u(i) - m_\alpha(i))^2}{m_\alpha(i)} \qquad (3)$$

$$\Delta\chi^2 = \chi_\gamma^2 - \chi_\alpha^2$$

Naturally, 0 was chosen as the threshold, and the result is illustrated in Fig. 3.

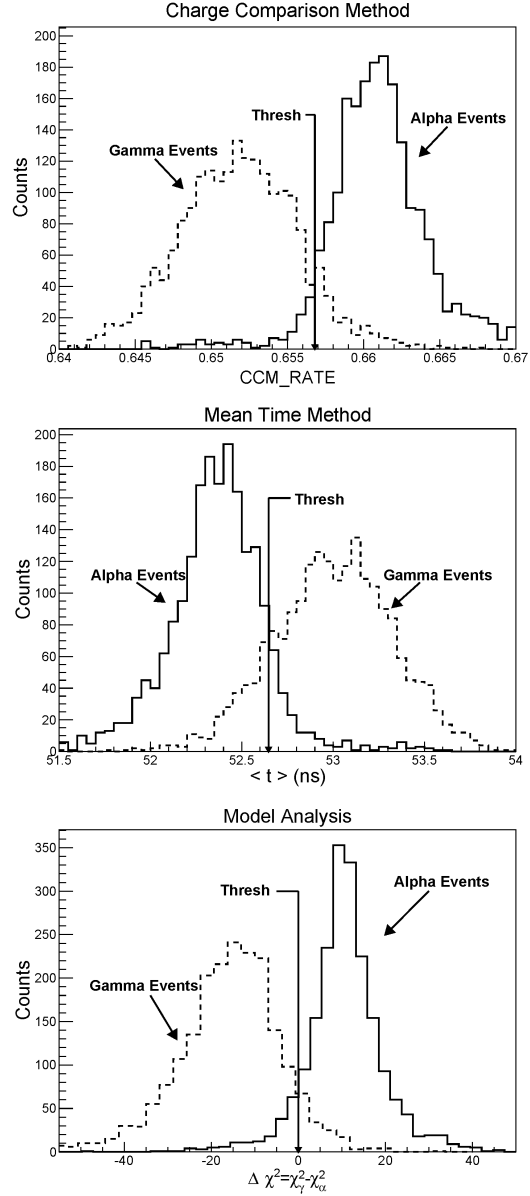

Fig. 3. Exact distributions of the feature value for each method, CCM, MTM, and GAMA. The solid lines correspond to alpha events, while the dotted ones correspond to gamma events.

### 3.4 Discrimination efficiency of PSD methods

To evaluate the separation of the alpha and gamma events, one intuitive approach commonly used in practice is to choose a threshold according to the distribution of feature values and then evaluate the percentage of properly discriminated events for each method. As shown in Fig. 3, a threshold, the intersection of the two peaks, is chosen to minimize the total incorrect rejection of both α and γ events. Hence, the discrimination efficiency can be calculated according to the corresponding histogram.

Meanwhile, another more quantitative approach is to use the following figure-of-merit (FOM) to define the separation [2]:

$$FOM = \frac{Peak\ separation}{FWHM_\alpha + FWHM_\gamma} \quad (4)$$

The separation using these three methods is shown in table 1. Defining $\eta_\alpha, \eta_\gamma$ as the discrimination efficiency, it can be found that the CCM and GAMA methods both achieve good discrimination efficiency.

TABLE I
EFFICIENCY COMPARISON OF DIFFERENT METHODS

| Method | Efficiency | | FOM |
|---|---|---|---|
| | $\eta_\alpha$ (%) | $\eta_\gamma$ (%) | |
| CCM | 93.9 | 90.8 | 0.686 |
| MTM | 85.2 | 85.9 | 0.503 |
| GAMA | 91.3 | 93.2 | 0.623 |

In addition, with further analysis of the pulse shape data of different energies, it can be found that the discrimination feature value and its statistical distribution vary with energy. Therefore, further study was undertaken to find the correlation between the pulse shape feature value and energy.

## 4  PSD Feature Value vs. Energy for the CCM

In this section, the CCM feature value and energy were plotted as a bi-parametric distribution 2D plot (see Fig. 4) to determine how the discrimination is influenced by energy.

The events are divided into different energy bins, with a 100 keV bin width. For gamma events between 300~1500 keV and alpha events between 1800~2500 keV separately, a Gaussian fit can be applied for each energy bin, to calculate the mean value and standard deviation (σ) of the distribution of the CCM feature value.

To analyse the correlation between the CCM feature value and the energy, a tentative linear fitting was used separately for alpha (1800~2500 keV) and gamma (300~1500 keV) (see Fig. 5). And it is verified using experimental data on higher-energy gamma events.

$$CCM = p_0 + p_1 \times Energy \quad (5)$$

According to Eq.1, the uncertainty of CCM can be expressed by the following equation, Where cov[$Q_p, Q_t$] means the covariance of $Q_p$ and $Q_t$.
:

$$\sigma_{CCM}^2 = \left(\frac{\partial CCM}{\partial Q_p}\right)^2 \sigma_{Q_p}^2 + \left(\frac{\partial CCM}{\partial Q_t}\right)^2 \sigma_{Q_t}^2 + 2\frac{\partial CCM}{\partial Q_p}\frac{\partial CCM}{\partial Q_t} \text{cov}\left[Q_p, Q_t\right] (6)$$

Considering the calculation of $Q_p$ and $Q_t$, the integral time base is $t_{MAX}$ ($t_{MAX}$ is aligned at 40ns for all pulses). For the chosen time interval (25.2~68 ns), the integral range before and after $t_{MAX}$ (40ns) are respectively 14.8ns and 28ns. Similarly $Q_p$ and $Q_t$ for this study is calculated as follows:

$$Q_p = \int_{t_{\max}-14.8ns}^{t_{\max}+28ns} i(t)\,dt = \int_{t_{\max}-14.8ns}^{t_{\max}+28ns} \left(i_0(t) + i_n(t)\right) dt \quad (7)$$

$$Q_t = \int_{t_{max}-40ns}^{t_{max}+120ns} i(t)\,dt = \int_{t_{max}-40ns}^{t_{max}+120ns} \left(i_0(t)+i_n(t)\right) dt \quad (8)$$

where $t_{max}$ is the time of the pulse maximum, $i_0(t)$ is the intrinsic signal without electronic noise, and $i_n(t)$ is the electronic noise.

From Eq.7 and Eq.8, it can be seen that the uncertainty of $Q_p$ and $Q_t$ can be separated into three almost independent parts:

$$\sigma_{Q_p}^2 = \sigma_{Q_p\_Jitter}^2 + \sigma_{Q_p\_Intrinsic}^2 + \sigma_{Q_p\_Noise}^2 \quad (9)$$

$$\sigma_{Q_t}^2 = \sigma_{Q_t\_Jitter}^2 + \sigma_{Q_t\_Intrinsic}^2 + \sigma_{Q_t\_Noise}^2 \quad (10)$$

where $\sigma_{Jitter}^2$ is the uncertainty caused by the variation of the time intervals, which results from the time jitter of the waveform alignment. For a given time interval, $\sigma_{Intrinsic}^2$ is the uncertainty caused by the intrinsic statistical fluctuation, and $\sigma_{Noise}^2$ is the uncertainty contributed by the electronic noise.

Thus, according to Eq.6, the uncertainty of CCM can also be separated into three parts:

$$\sigma_{CCM}^2 = \sigma_{CCM\_Jitter}^2 + \sigma_{CCM\_Intrinsic}^2 + \sigma_{CCM\_Noise}^2 \quad (11)$$

First, the $\sigma_{CCM\_Jitter}^2$, the uncertainty in CCM caused by time jitter, can be similarly described as follows:

$$\sigma_{CCM\_Jitter} \approx \frac{|i(t_2)-i(t_1)|}{\int_{total} i(t)\,dt} \cdot \sigma_{t_{max}} \quad (12)$$

Where $i(t_1)$ and $i(t_2)$ are the amplitudes at the boundaries of integral time interval, and $\sigma_{tmax}$ is the time jitter of the pulse alignment. That is to say, the uncertainty can be minimized by choosing the proper integral time interval, making the amplitude values at the two boundaries similar. Moreover, a Finite Impulse Response (FIR) digital filter can be used to reduce the time jitter.

In conclusion, with proper pulse processing, the uncertainty cause by time jitter can be negligible compared with other causes.

For the scintillation detector, the energy resolution is similarly proportional to $1/\sqrt{E}$. Both $Q_t$ and $Q_p$ are estimates of the deposited energy. It is reasonable to assume that

$$\frac{\sigma_{Q_p\_Intrinsic}}{Q_p}, \frac{\sigma_{Q_t\_Intrinsic}}{Q_t} \propto \frac{1}{\sqrt{E}} \quad (13)$$

Then, the uncertainty of CCM caused by the intrinsic statistical fluctuation can be expressed as follows:

$$\sigma_{CCM\_Intrinsic}^2 \propto \frac{1}{E} \quad (14)$$

Similarly, for the electronic noise, the uncertainty can be simplified as

$$\sigma^2_{CCM\_Noise} \propto \frac{1}{E^2} \qquad (15)$$

Finally, the uncertainty of CCM in Eq.11 can be represented as

$$\begin{aligned}\sigma_{CCM} &= \sqrt{\sigma^2_{CCM\_Jitter} + \sigma^2_{CCM\_Intrinsic} + \sigma^2_{CCM\_Noise}} \\ &\approx \sqrt{\left(\frac{c1}{E}\right)^2 + \left(\frac{c2}{\sqrt{E}}\right)^2}\end{aligned} \qquad (16)$$

Finally, the CCM feature value and its σ were separately fitted with the experimental data in the lead shield, using Eq.5 and Eq.16. The result is plotted in Fig. 4 and Fig. 5:

$$\begin{aligned} CCM_\gamma &= 0.7346 - 3.572 \times 10^{-6} E \\ \sigma_{CCM_\gamma} &= \sqrt{\left(\frac{1.148}{E}\right)^2 + \left(\frac{0.081}{\sqrt{E}}\right)^2} \\ CCM_\alpha &= 0.7434 - 2.674 \times 10^{-6} E \\ \sigma_{CCM_\alpha} &\approx \frac{0.101}{\sqrt{E}} \end{aligned} \qquad (17)$$

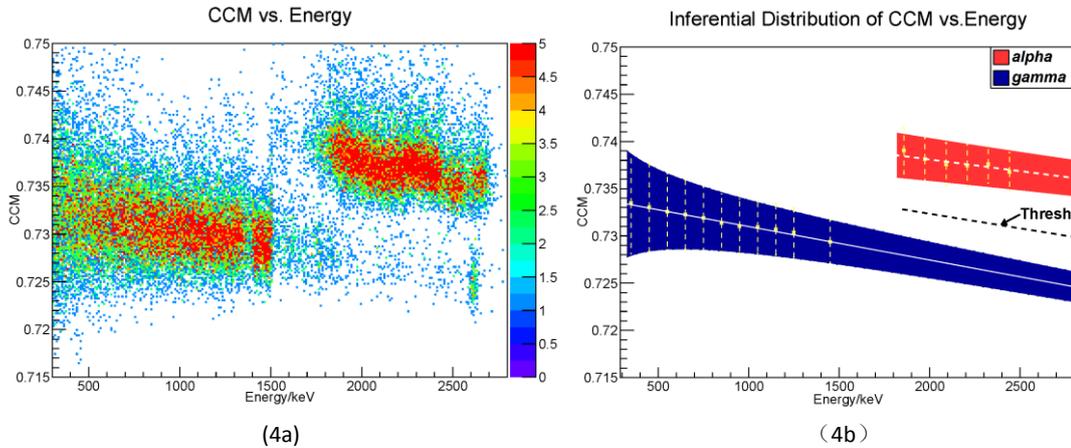

(4a)　　　　　　　　　　　　　　　　（4b）

Fig. 4.　4a is the measured CCM distribution vs. energy, while 4b is the calculated distribution using Eq.17, with the energy-dependent CCM feature value and its sigma broadening. The threshold line is calculated using the interaction of alpha and gamma.

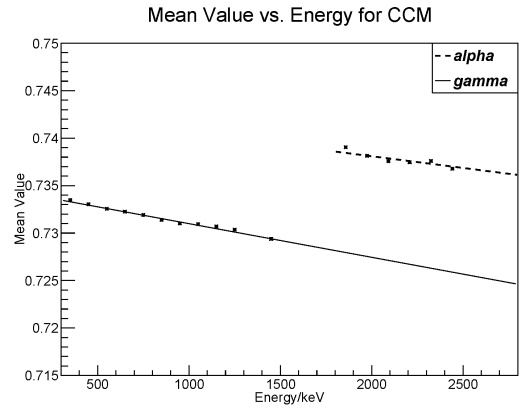

(5a)

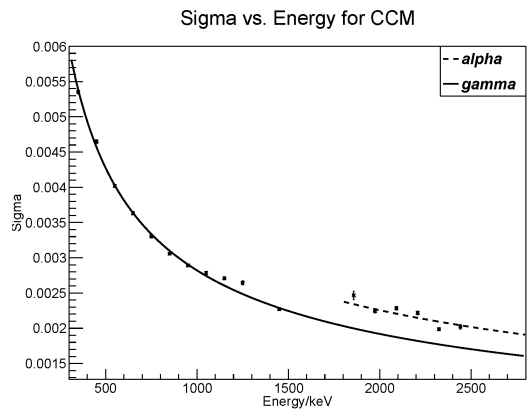

(5b)

Fig. 5.  5a shows how the CCM feature value (mean) of gamma and alpha varies with energy. 5b shows how the sigma of the CCM feature value varies with energy. The fitting lines are calculated using Eq.17.

Clearly, with fitting Eq.17 and Fig. 4, an optimized and energy-dependent threshold line can be drawn on the plot, rather than a single fixed threshold.

Moreover, although the above data fitting was performed using gamma events between 300~1500 keV, it can be seen in Fig. 4 that there are a number of gamma events above 1600 keV. For verification, a separation of gamma and alpha events was performed using the energy-dependent threshold line. As shown in Fig 6, from the separated gamma spectrum, the photon peaks of $^{208}$Tl (2615keV) and $^{214}$Bi (1765keV) are clearly visible.

For a further verification of Eq.17, all events in the energy bin $2615 \pm 100$ keV were plotted as a distribution of the CCM feature value, as shown in Fig. 7. Two Gaussian peaks can be clearly seen, and Gaussian fitting was applied to the lower peak of the gamma distribution. The fitted value and σ are 0.7251 ± 0.0016, which is in very good accordance with the result calculated using Eq.17, i.e., 0.7253 ± 0.0016.

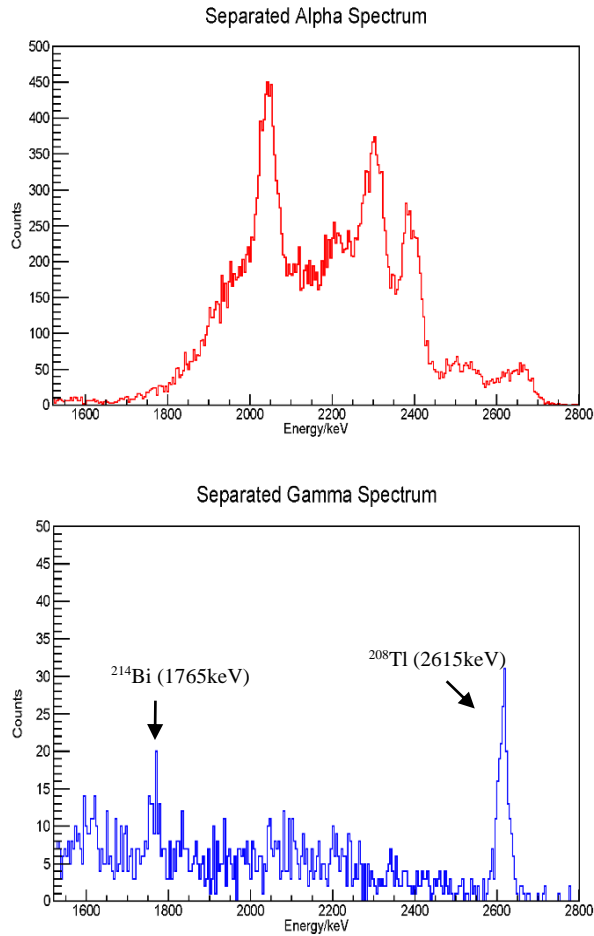

Fig. 6. Separated gamma spectrum with energy-dependent threshold. Photon peaks of $^{214}$Bi and $^{208}$Tl are visible.

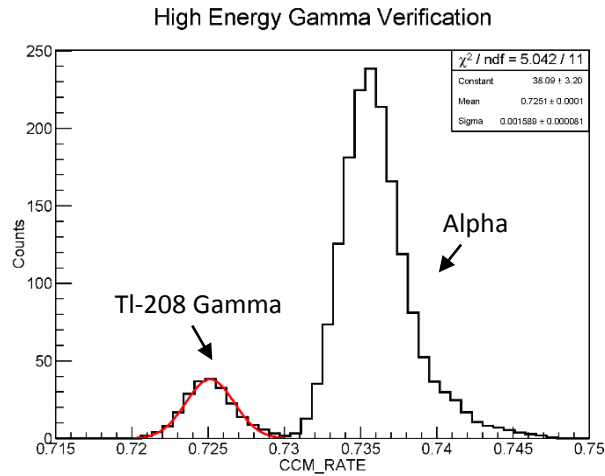

Fig. 7. Distribution of CCM feature value, with events in energy region 2615 ± 100 keV. Two Gaussian peaks of gamma ($^{208}$Tl 2615 keV photon peak) and alpha (intrinsic radioactivity) events are clearly visible.

## 5  Application and Discussion

To verify the validity of the fitting equation (Eq.17) inferred in part 4, a new data set was taken with the lead shielding removed (the spectrum measured is shown as Fig. 8).

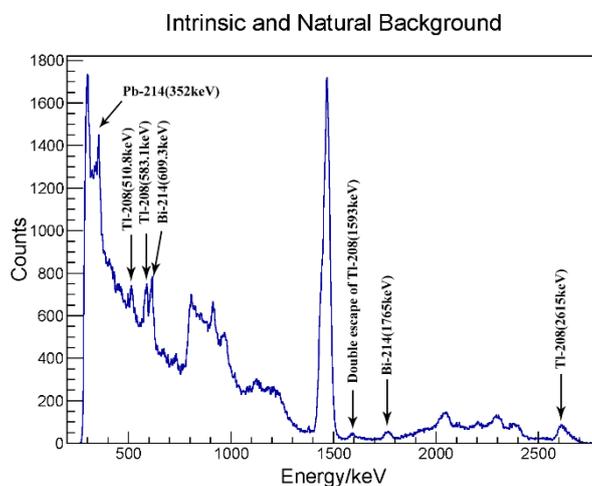

Fig. 8.  Origin spectra of intrinsic and natural background. Some characteristic peaks of nuclide $^{208}$Tl and $^{214}$Bi can be seen.

As expected, the existence of $^{208}$Tl and $^{214}$Bi can be confirmed. However, in the origin spectra before PSD, the high-energy gamma peaks are nearly submerged in the intrinsic alpha background (Fig. 9).

With the PSD method and threshold optimized using Eq.17, however, as shown in Fig. 9, the photon peak at 2615 keV and its single escape and double escape peaks of $^{208}$Tl (from the decay chain of $^{232}$Th), as well as the 1764.5 keV photon peak of $^{214}$Bi (from the decay chain of $^{222}$Rn), can be clearly observed in the discriminated gamma spectrum.

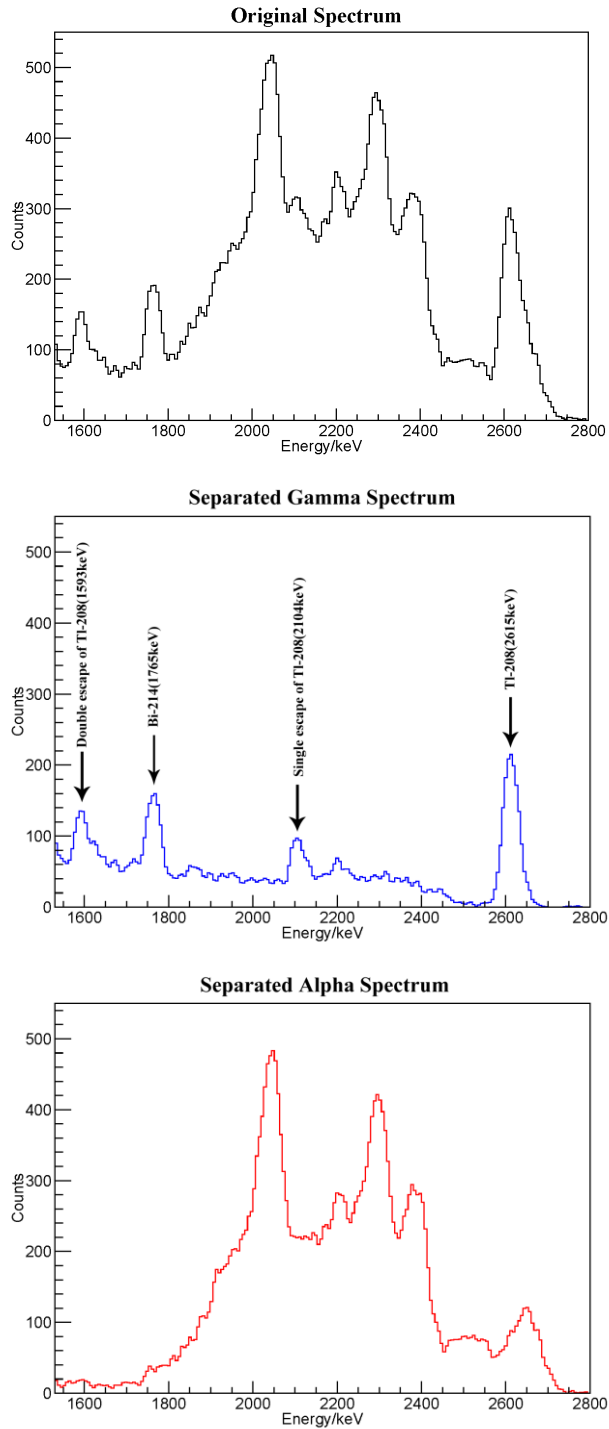

Fig. 9. The CCM-discriminated spectra (energy region between 1520 keV and 2800 keV). 9a: original spectrum (before CCM PSD). 9b and 9c: alpha and gamma spectra with CCM PSD. A suitable energy-dependent threshold is chosen according to the CCM distribution using Eq.17.

Meanwhile, other characteristic peaks of $^{208}$Tl and $^{214}$Bi can also be found in the lower-energy region, such as 510.8 keV and 583.1 keV for $^{208}$Tl and 609 keV for $^{214}$Bi (see Fig. 8).

It was also shown in these experiments that the distribution of the CCM is related to the environmental temperature, which should be carefully studied later.

# 6 Conclusion

In this paper, the charge comparison method for LaBr$_3$:Ce is quantitatively analysed and compared with two other typical PSD methods, using digital pulse processing. It is shown that the CCM and GAMA methods are both applicable in γ-α discrimination. The correlation between the CCM feature value distribution and the radiation energy is studied in detail, and a fitting equation is inferred and verified using high-energy γ experimental data. Although the reason for the pulse shape difference in LaBr$_3$:Ce remains unclear, it was found that the energy-dependent CCM feature value can be fitted very well with a linear equation (i.e., Eq.17).

Meanwhile, based on the resulting fitting equation, an optimum energy-dependent threshold line for the PSD could be given, which has been shown to be valuable for low-activity high-energy γ measurement by suppressing the alpha background. These results support the need for further study of the application to radioactivity measurement, as well as identification of the reason for the pulse shape difference in LaBr$_3$:Ce.

## Acknowledgements

This work was supported by the National Natural Science Foundation of China (No. 11305093 & No.11175099) and the Tsinghua University Initiative Scientific Research Program (No. 2011Z07131 & No. 2014Z21016).